\documentclass[11pt]{article}
\usepackage[symbol]{footmisc}
\def\correspondingauthor{\footnote{Corresponding author.  }}
\usepackage[left=2cm,top=2cm,right=2cm,bottom=2cm]{geometry}
\usepackage{amsmath, amssymb}

\usepackage{graphicx}
\usepackage{graphics}
\usepackage{epstopdf}
\usepackage{subfigure}
\usepackage{amsfonts}
\usepackage{sectsty}
\usepackage{sectsty}
\usepackage{hyperref}
\usepackage{cite}
\begin{document}
	\begin{center}
	\large{\bf{Gravitational lensing effect in traversable wormholes}} \\
	\vspace{5mm}
\normalsize{Nisha Godani$^1$ and Gauranga C. Samanta$^{2}{}$\correspondingauthor{} }\\
\normalsize{$^1$Department of Mathematics, Institute of Applied Sciences and Humanities\\ GLA University, Mathura, Uttar Pradesh, India\\
$^2$P. G. Department of Mathematics,
Fakir Mohan University, Balasore, Odisha, India}\\
\normalsize {nishagodani.dei@gmail.com\\gauranga81@gmail.com}
\end{center}

\begin{abstract}
The present paper is intended for studying the effect of strong gravitational lensing in the context of charged wormhole. To study this effect, the conditions determining the existence of photon spheres at and outside the throat are obtained. The necessary and sufficient conditions
for the existence of photon spheres at or outside the throat of the charged wormhole is derived. Furthermore, photon spheres are investigated in three cases for three different forms of redshift function. These three cases include the existence of effective photon spheres (i) at the throat, (ii) outside the throat and (iii) both at and outside the throat. Consequently, these provide the information about the formation of infinite number of concentric rings and may lead to the detection of wormhole geometries.   		
\end{abstract}

\textbf{Keywords:}  Wormhole; Gravitational lensing; Photon spheres.

\section{Introduction}

Gravitational lensing, first identified in the deflection of light by sun, is a significant and effective phenomenon for the detection of astrophysical objects such as black holes. It has attracted the attention of researchers and opened a vast area of research. Earlier, the research was carried out on the gravitational lensing in the presence of weak field but it was not possible to differentiate various asymptotically flat solutions in the weak
 field regime \cite{Schneider}. Then the research progressed in the strong field regime \cite{Frittelli,Virbhadra}, where photon and antiphoton spheres have a significant role. Photon spheres are determined by radii at which photons have unstable circular orbit. However, for antiphoton spheres, orbits are stable. Photons are captured by photon sphere where they rotate multiple times until  escaped to infinity by a small perturbation. In the strong gravitational field, by an analytic approximation method, it was shown that the deflection angle diverges at the photon sphere and this divergence is logarithmic \cite{Bozza}. It has also been proved to be generic feature with respect to a spherical, static and symmetric metric of space-time \cite{Bozza1}.   The concept of strong lensing has been applied for the investigation of black holes \cite{Bhadra, Bozza2, Vasquez} which has increased the domain of information about their signatures.

Further, wormhole is one of the most important topics in the past few decades. It is a region of space-time that provides a shorter route for the connection of two far objects. These are supported by matter not obeying the energy conditions at the throat in the framework of general relativity. The idea of wormhole was first suggested by Flamm \cite{flamm} after the discovery of Schwarzschild's black hole solutions. Then a bridge like structure was proposed by Einstein and Rosen known as Einstein-Rosen bridge \cite{eins-ros}.  The interest in traversable wormholes has been developed by the study of Morris, Thorne and Yurtsever \cite{morris1,Morris}. Although, the construction of wormhole solutions with ordinary matter is a challenging problem, nevertheless, this problem has been resolved in the framework of GR for rotating cylindrical wormholes, and wormhole solutions satisfying weak energy condition  have been investigated in \cite{Bronnikov1, Bronnikov2, Bronnikov3}.  The same problem has also been taken into account by many researchers using modified theories of gravity such as $f(R)$ gravity, $f(R,T)$ gravity \cite{saiedi, najafi, Rahaman1, Samanta19, Godani19, Samantaepjc, ng_chinese,  Zubair1, Elizalde51, Sharif3}.
Gravitational lensing in wormholes has been explored in many aspects and compared with the study of black holes \cite{Nandi,Rahaman,Dey,Bhattacharya,Abe,Nakajima,Sharif,Tsukamoto1,Tsukamoto2, Nandi1, Nandi2}.
It is examined for wormholes with negative masses in \cite{Cramer, Safonova}. Morris Thorne wormholes are also obtained to act as convergent lenses \cite{Tejeiro}.
 By considering two conditions (i) when the source and the observer are on the same side and (ii) when they are on the different sides, the presence of photon and antiphoton spheres on both sides of the throat is proved and throat is found to act as a photon sphere in spherically symmetric wormhole  \cite{Shaikh}. Subsequently, Bronnikov and Baleevskikh \cite{Bronnikov2019gc} discussed, if the wormhole is symmetric, then the photon sphere is necessarily located on the throat. The necessary and sufficient conditions for the existence of photon spheres at or outside the throat of wormhole are obtained in \cite{Shaikh2}. Falco et al.\cite{Falco2020}  derive the equations of motion of a test particle in the equatorial plane around a static and spherically symmetric wormhole influenced by a radiation field including the general relativistic Poynting-Robertson effect, and the authors find out only a practical implication in diagnosing a BH from a WH, but also other interesting applications. Subsequently, the same authors\cite{Falco2021}  
 develop a model-independent procedure to single out static and spherically symmetric wormhole solutions based on the general relativistic Poynting-Robertson effect and the extension of the ray-tracing formalism in generic static and spherically symmetric wormhole metrics.
 The study of gravitational lensing is also carried out within the context of alternative theories of gravity. The effect of $f(R,T)$ theory of gravity on gravitational lensing is determined and the results are compared with the corresponding results in Einstein's general theory of relativity \cite{Alhamzawi}. The deflection angle of light for charged wormholes is calculated using Einstein-Maxwell-dilaton theory \cite{Jusufi}. Scalar tensor theory is also used to study the effect of lensing in wormholes and analyze its relation with energy conditions \cite{Shaikh1}.

In this paper, we have considered  charged wormholes which are also studied in literature using the concepts of  scalar field, $f(R,T)$ gravity etc. \cite{Kim, Gonjalez, Sharif1, Moraes4}. The motivation of the present work is to examine the effect of gravitational lensing on charged wormholes by deriving the necessary and sufficient conditions for the existence of photon spheres at or outside the throat of the charged wormhole. Subsequently, the existence of photon spheres is examined for three different and novel forms of redshift function and the formation of infinite number of  concentric rings is obtained. Furthermore, we have used the background of $f(R,T)$ gravity for charged wormholes and examined the validity of null and weak energy conditions, which successively leads to determine the type of matter threading for the geometry of the charged wormholes.

\section{Charged wormholes and photon spheres}
The spherically symmetric  charged wormhole metric is given by
\begin{equation}\label{charge}
ds^2=-e^{2\Phi_{eff}(r)}dt^2+\frac{dr^2}{1-\frac{b_{eff}(r)}{r}}+r^2(d\theta^2+\sin^2\theta d\phi^2),
\end{equation}
where $(t, r, \theta, \phi)$ are the usual space-time coordinates. The angles $\theta$ and $\phi$ vary from 0 to $\pi$ and 0 to $2\pi$, respectively.  The functions $\Phi_{eff}(r)$ and $b_{eff}(r)$ represent effective redshift and shape functions respectively. The redshift function $\Phi_{eff}(r)$ is responsible for  determining gravitational redshift. The function $b_{eff}(r)=b(r)-\frac{q}{r}$, where $b(r)$ denotes the shape function effecting the shape of the wormhole and $q$ acts as a free parameter. The radial coordinate $r$ varies from the minimum value $r_0>0$ to infinity. The minimum value $r_0$ is called the  throat
of the wormhole. The shape function should satisfy the following conditions in the context of charged wormholes: (i) $\frac{b(r)}{r}-\frac{q^2}{r^2}<1$ for $r>r_0$,  (ii) $b(r_0)=r_0\Big(1+\frac{q^2}{r_0^2}\Big)$ at $r=r_0$, (iii) $\frac{b(r)}{r}-\frac{q^2}{r^2}\rightarrow 0$ as $r\rightarrow \infty$, (iv) $b'(r_0)+\frac{q^2}{r_0^2}<1$ for $r>r_0$.
The presence of horizons must be avoided in traversable wormholes. To avoid horizons, we must have $e^{2\Phi_{eff}(r)}\neq 0$. For asymptotically flat wormholes, $e^{2\Phi_{eff}(r)}\rightarrow 1$ as $r \rightarrow \infty$.

For metric \eqref{charge}, we restrict to $\theta = \frac{\pi}{2}$ and the same results can be applied for all values of $\theta$. Let $\mathfrak{L}$ denote the Lagrangian for the motion of a photon. Then with respect to metric \eqref{charge}, we have \begin{equation}\label{L}
2\mathfrak{L}=-e^{2\Phi_{eff}(r)}\dot{t}^2+\left(1-\frac{b_{eff}(r)}{r}\right)^{-1}\dot{r}^2+r^2\dot{\phi}^2,
\end{equation}
where dot denotes the differentiation with respect to the affine parameter.

Since the Lagrangian does not depend on the variables $t$ and $\phi$, therefore

\begin{equation}\label{}
\frac{\partial \mathfrak{L}}{\partial \dot{t}}=-e^{2\Phi_{eff}(r)}\dot{t}=-E
\end{equation}
and
\begin{equation}\label{}
\frac{\partial \mathfrak{L}}{\partial \dot{\phi}}=r^2\dot{\phi}=L,
\end{equation}
where $E$ and $L$ stand for the energy and angular momentum of the photon respectively. By the null geodesic condition, $g_{\mu\nu}\dot{x}^{\mu}\dot{x}^{\nu}=0$, we have
\begin{equation}\label{geodesic}
E^2=e^{2\Phi_{eff}(r)}\left(1-\frac{b_{eff}(r)}{r}\right)^{-1}\dot{r}^2+V_{eff},
\end{equation}
where $V_{eff}$ is the effective potential defined as
\begin{equation}\label{V}
V_{eff}=L^2e^{2\Phi_{eff}(r)}r^{-2}.
\end{equation}
This gives the deflection angle for the metric \eqref{charge} as \cite{Bozza}

\begin{equation}\label{alpha}
\alpha=2\int_{r_{tp}}^{\infty}\frac{\left(1-\frac{b_{eff}(r)}{r}\right)^{-\frac{1}{2}}dr}{r\sqrt{\frac{r^2}{u^2e^{2\Phi_{eff}(r)}}-1}}-\pi,
\end{equation}
where $u$ is called the impact parameter and $r_{tp}$ is known as the turning point given by $\frac{dr}{d\phi}=0$. The impact parameter is defined as $u=\frac{L}{E}$ and the relation between $u$ and $r_{tp}$ is
\begin{equation}\label{}
u=r_{tp}e^{-\Phi_{eff}(r)}.
\end{equation}
The deflection angle has logarithmic divergence to infinity as the turning point approaches to a photon sphere which corresponds to an ustable orbit which satisfy the following conditions: $\dot{r}=0$, $\ddot{r}=0$ and $\dddot{r}>0$. These conditions can be written in  terms of effective potential as
\begin{equation}\label{condition1}
V_{eff}(r_{ph})=E^2,  \frac{dV_{eff}(r_{ph})}{dr}=0 \text{ and }   \frac{d^2V_{eff}(r_{ph})}{dr^2}<0,
\end{equation}
where $r_{ph}$ is the radius of the photon sphere. By \eqref{condition1}, $V_{eff}$ is maximum at $r=r_{ph}$. The effective potential may possess an extreme value at $r=r_0$, i.e. the throat of wormhole may  act as a photon sphere. For metric \eqref{charge}, the conditions \eqref{condition1} reduce to

\begin{equation}\label{condition2a}
u_{ph}=r_{ph}e^{-\Phi_{eff}(r_{ph})}
\end{equation}

\begin{equation}\label{condition2b}
 \Phi^{'}_{eff}(r_{ph})=\frac{1}{r_{ph}}
\end{equation}
and
\begin{equation}\label{condition2c}
\Phi^{''}_{eff}(r_{ph}) < -\frac{1}{r_{ph}^2}.
\end{equation}

Rewriting Eq. \eqref{geodesic},
\begin{equation}\label{geodesic1}
\dot{r}^2+e^{-2\Phi_{eff}(r)}\left(1-\frac{b_{eff}(r)}{r}\right)(V_{eff}-E^2)=0.
\end{equation}

At throat, $b(r_0)=r_0\Big(1+\frac{q^2}{r_0^2}\Big)$, which implies $\dot{r}=0$.
The throat can act as an effective photon sphere, if
\begin{equation}\label{condition3}
\ddot{r}=0 \text{ and } \dddot{r}>0.
\end{equation}

The proper radial coordinate is given by
\begin{equation}\label{proper}
l(r)=\pm\int_{r_0}^{r}\frac{dr}{\sqrt{1-\frac{b_{eff}(r)}{r}}},
\end{equation}
where $l(r_0)=0$
Using the flare out condition $b'(r_0)+\frac{q^2}{r_0^2}<1$, the conditions \eqref{condition3} can be written in terms of $V_{eff}$ as
\begin{equation}\label{condition4}
V_{eff}(r_0)=E^2 \text{ and } \frac{dV_{eff}(r_{0})}{dr}<0.
\end{equation}
Further, the conditions \eqref{condition4}  take the form

\begin{equation}
u_{0}=r_{0}e^{-\Phi_{eff}(r_{0})}
\end{equation}
and
\begin{equation}
\Phi^{'}_{eff}(r_{0})<\frac{1}{r_{0}}.
\end{equation}

In terms of proper radial coordinates, Eq. \eqref{geodesic1} reduce to
\begin{equation}
e^{2\Phi_{eff}(r(l))}\dot{l}^2+V_{eff}=E^2,
\end{equation}
where $V_{eff}=\frac{L^2e^{2\Phi_{eff}(r(l))}}{r^2(l)}$.
\\

At the throat,
\begin{equation}\label{l}
V_{eff}(r_0)=E^2, \frac{dV_{eff}}{dl}\Big|_{r_0}=0 \text{ and} \frac{d^2V_{eff}}{dl^2}\Big|_{r_0}<0.
\end{equation}
Conditions \eqref{l} in terms of proper radial coordinate $l$ at $r=r_0$ are similar to the conditions \eqref{condition1} in terms of the radial coordinates $r$ at $r=r_{ph}$.

When the effective photon spheres are obtained at the throat of wormhole and outside the throat of wormhole, then we have two sets of infinite number of relativistic images formed
due to the strong gravitational lensing of light. In this case, it is clear that the maxima occurs at both $r=r_0$ and $r=r_{ph}$. The value of $V_{eff}$ corresponding to the photon sphere at the throat must be greater than the value of $V_{eff}$   corresponding to the outer photon sphere. The necessary and sufficient conditions for the existence of photon spheres both at and outside the throat are

 \begin{equation}
 \Phi^{'}_{eff}(r_{0})<\frac{1}{r_{0}}
 \end{equation}
 and
 \begin{equation}
 \frac{e^{2\Phi_{eff}(r_{0})}}{r_0^2}>\frac{e^{2\Phi_{eff}(r_{ph})}}{r_{ph}^2}.
 \end{equation}

\section{Effect of strong gravitational lensing in charged wormholes}
 Strong gravitational lensing  is a physical phenomenon which can be used for the detection of wormhole structures. In the previous section, we found some conditions for the existence of photon spheres at the throat of wormhole or outside the throat of wormhole in the context of charged wormhole. These correspond to unstable photon orbits which consequently provide the formation of infinite number of Einstein's rings. The metric \eqref{charge} is dependent mainly on two functions: $\Phi(r)_{eff}$ and $b(r)$. In this section, we have restricted to $b(r)=\frac{r_0^2}{r}$ and three particular forms of function $\Phi_{eff}(r)$ in order to show the existence of photon spheres (i) at the throat, (ii) outside the throat and (iii) both at and outside the throat. Thus, to achieve this aim, we have divided this section into three cases with respect to three forms of  $\Phi_{eff}(r)$ which are as follow:\\

\noindent
\textbf{Case 1:} $\Phi_{eff}(r)=\frac{1}{2}\log_e(1+\frac{2r_0}{3r})$\\

\noindent
For this form of $\Phi_{eff}(r)$,  $V_{eff}/L^2=\frac{1}{r^2} + \frac{2r_0}{3r^3}$ from \eqref{V}. Then conditions \eqref{condition4} are satisfied at $r=r_0$. This shows the existence of photon sphere at the throat of wormhole. Using Eq. \eqref{proper}, we have $r^2=l^2+r_0^2$, which reduces to the effective potential as $V_{eff}/L^2=\frac{1}{l^2+r_0^2} + \frac{2r_0}{3(l^2+r_0^2)^{3/2}}$. This form of $V_{eff}$ gives $\frac{dV_{eff}}{dl}\Big|_{r_0}=0$ and $\frac{d^2V_{eff}}{dl^2}\Big|_{r_0}<0$, which means that $V_{eff}$ is maximum at $r=r_0$. Hence, conditions \eqref{l} are fulfilled. In Fig. 1(a), we have plotted $V\equiv\frac{V_{eff}}{L^2}$ with respect to the proper radial coordinate $l$ which attains a maxima at $l=0$, i.e. at the throat. Further, using \eqref{alpha}, we have plotted the deflection angle $\alpha$ with respect to impact parameter $u$ in Fig. 1(b). It is diverging logarithmically to infinity at the turning point and then it  decreases as $u$ increases. In this case,  the turning point  $r_{tp}=r_0$. In both Figures 1(a) and 1(b), $V$ and $\alpha$ are plotted, respectively,  which show the existence of effective photon sphere at the throat for three values of $r_0$.\\

\noindent
\textbf{Case 2:} $\Phi_{eff}(r)=\frac{1}{2}\log_e(1-\frac{2\beta}{3r})$, where $\beta>r_0$\\

\noindent
For this form of $\Phi_{eff}(r)$,  $V_{eff}/L^2=\frac{1}{r^2} - \frac{2\beta}{3r^3}$, where $\beta$ is any number greater then $r_0$. Then we have $\frac{dV_{eff}}{dr}=0$ and $\frac{d^2V_{eff}}{dr^2}<0$ at $r=\beta$. So, the conditions \eqref{condition1} are satisfied at $r=\beta$ and $V_{eff}$ has a maxima at this point. Thus, $r_{ph}=\beta$ is the radius of the photon sphere occurring outside the throat. Thus, this form of $\Phi(r)$ generates a photon sphere outside the throat of wormhole. In Fig. 2(a), $V\equiv\frac{V_{eff}}{L^2}$ is plotted with respect to the proper radial coordinate $l$ for three values of $\beta$. It attains maxima at $l=\pm\sqrt{\beta^2-r_0^2}$, which depicts the presence of photon sphere outside the throat. Further, the deflection angle $\alpha$ is calculated and plotted in Fig. 2(b) with respect to $u$ for three different values of $\beta$.
It is found to diverge logarithmically to infinity at the turning point  and then it decreases as $u$ increases.  Hence, we have a outer photon sphere in this case which  contributes in the formation of relativistic images only outside the throat.\\

\noindent
\textbf{Case 3:} $\Phi_{eff}(r)=\frac{1}{2}\log_e\left(1-\frac{2(\beta+\gamma)}{3r}+\frac{\beta\gamma}{r^2}\right)$, where $r_0<\beta<\gamma$.\\

\noindent
The aim for considering this form of $\Phi_{eff}(r)$ is to show the existence of effective photon spheres at the throat of wormhole as well as its outside. This requires that the conditions \eqref{condition4} or \eqref{l} must be satisfied at $r=r_0$  and conditions \eqref{condition1} must be fulfilled outside the throat corresponding to some radius of photon sphere. Between these two values of $r$, $V_{eff}$ must possess a minimum value. For this case, we have   $V\equiv V_{eff}/L^2=\frac{1}{r^2} - \frac{2(\beta+\gamma)}{3r^3}+\frac{\beta\gamma}{2r^4}$. $V_{eff}$ has a maxima at $r=r_0$ and $r=\gamma$. It possesses a minima at $r=\beta$. In Figures 3(a)-3(e), we have plotted $V$ with respect to the proper radial coordinate $l$ for different values of $\beta$ and $\gamma$. In Figures 3(a) and 3(b), only the wormhole throat is acting as an effective photon sphere.
In Figures 3(c) and 3(d), there are existing two photon spheres. One is effective photon sphere  at the throat and  other is outer photon sphere. The value of $V_{eff}$ is maximum at the throat in comparison of its value corresponding to outer photon sphere. This means that the throat will participate in the formation of relativistic images due to strong gravitational lensing. In Fig. 3(e), the value of $V_{eff}$ is lesser than its value corresponding to outer photon sphere. This shows that the throat will not participate in the formation of relativistic images due to strong gravitational lensing for chosen values of $\beta$ and $\gamma$. Further, in Fig. 3(f), we have plotted $V_{eff}$ with respect to $r$ for $\beta=3$, $\gamma=5, 5.5, 6$ and $r_0=2.3$. It has maxima at the throat with higher height than the maxima outside. There is a minima between two maxima. Furthermore, we have calculated deflection angle and plotted with respect to impact parameter $u$. In Figures 4(a) and 4(b), $\alpha$ diverges only at the throat. In Figures 4(c) and 4(d), $\alpha$ diverges at two points corresponding to the radii of the throat and  outer sphere. After diverging at the throat, first it decreases, then increases for outer sphere and after that it again decreases.  In Fig. 3(e), since only outer sphere is taking part in the formation of rings, $\alpha$ will diverge for outer sphere. In Fig. 4(e), we can see the diverging nature of $\alpha$ corresponding to outer sphere.
Thus, the same set of values of $\beta$ and $\gamma$ is taken in Figures 3(a)-3(e) and  Figures 4(a)-4(e) to explain the nature of photon spheres by effective potential and deflection angle respectively.
\begin{figure}
	\centering
	\subfigure[ This is plot for effective potential $V\equiv \frac{V_{eff}}{L^2}$ with respect to proper radial coordinate $l$. It shows the maximum value of $V$ at the throat of wormhole.]{\includegraphics[height=6cm, width=7cm]{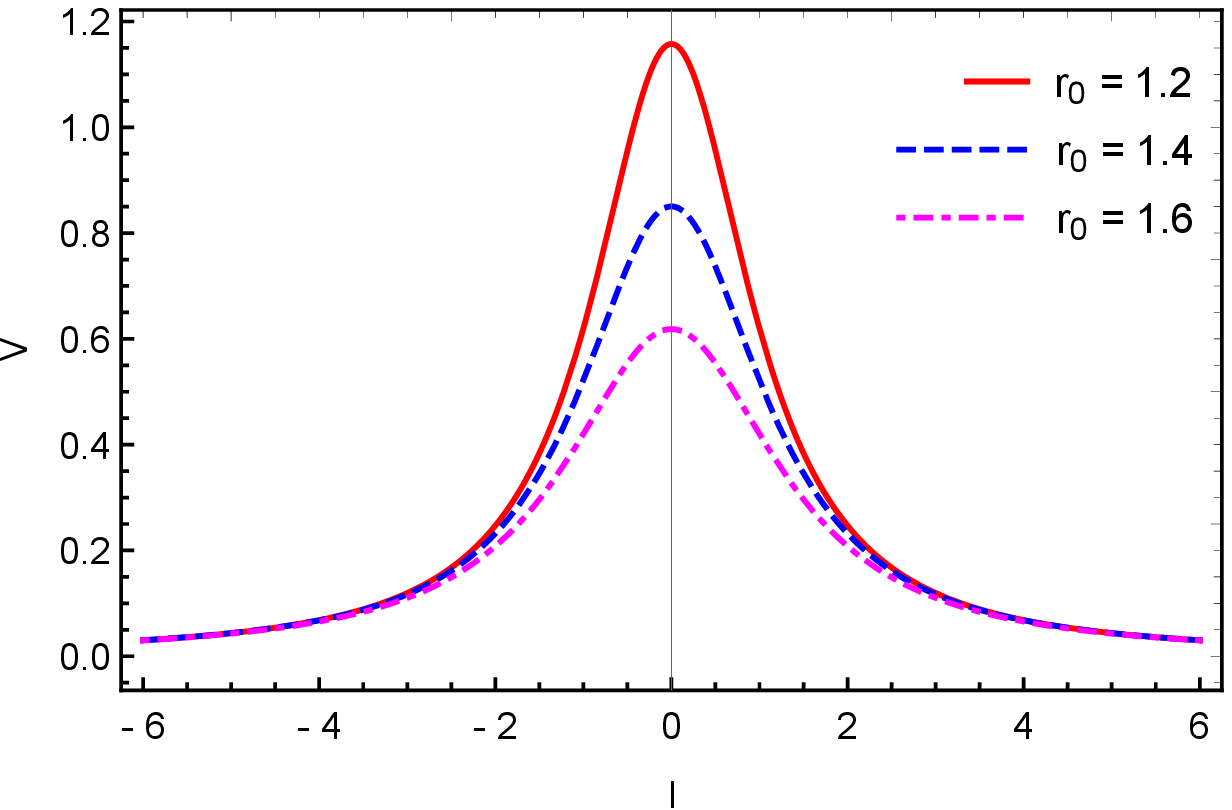}}\hspace{.5cm}
	\subfigure[This is plot for deflection angle $\alpha$ with respect to impact parameter $u$. It shows the divergence of $\alpha$ corresponding to the radius of the throat.]{\includegraphics[height=6cm, width=7cm]{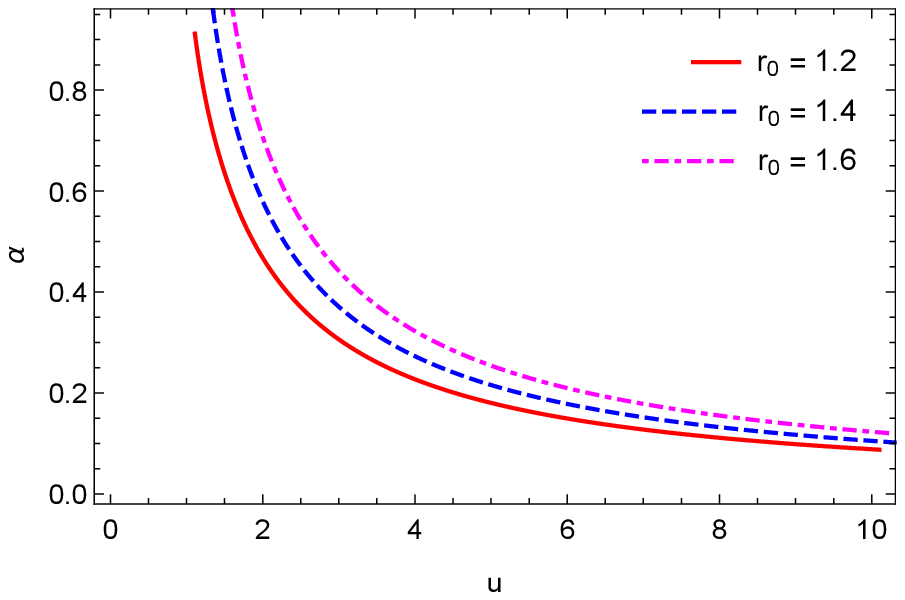}}\hspace{.3cm}
	\caption{Plots for effective potential and deflection angle for redshift function $\Phi_{eff}(r)=\frac{1}{2}\log_e(1+\frac{2r_0}{3r})$.}
\end{figure}

\begin{figure}
	\centering
	\subfigure[ This is plot for effective potential $V\equiv \frac{V_{eff}}{L^2}$ with respect to proper radial coordinate $l$. It shows the maximum value of $V$ corresponding to the radius of photon sphere occuring outside the throat.]{\includegraphics[height=6cm, width=7cm]{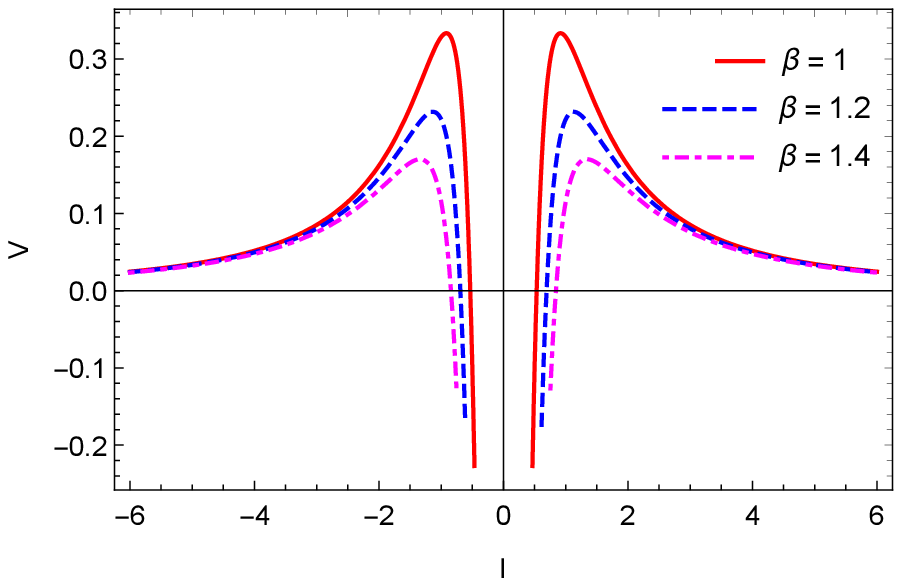}}\hspace{.5cm}
	\subfigure[This is plot for deflection angle $\alpha$ with respect to impact parameter $u$. It shows the divergence of $\alpha$ corresponding to the radius of photon sphere occuring outside the throat.]{\includegraphics[height=6cm, width=7cm]{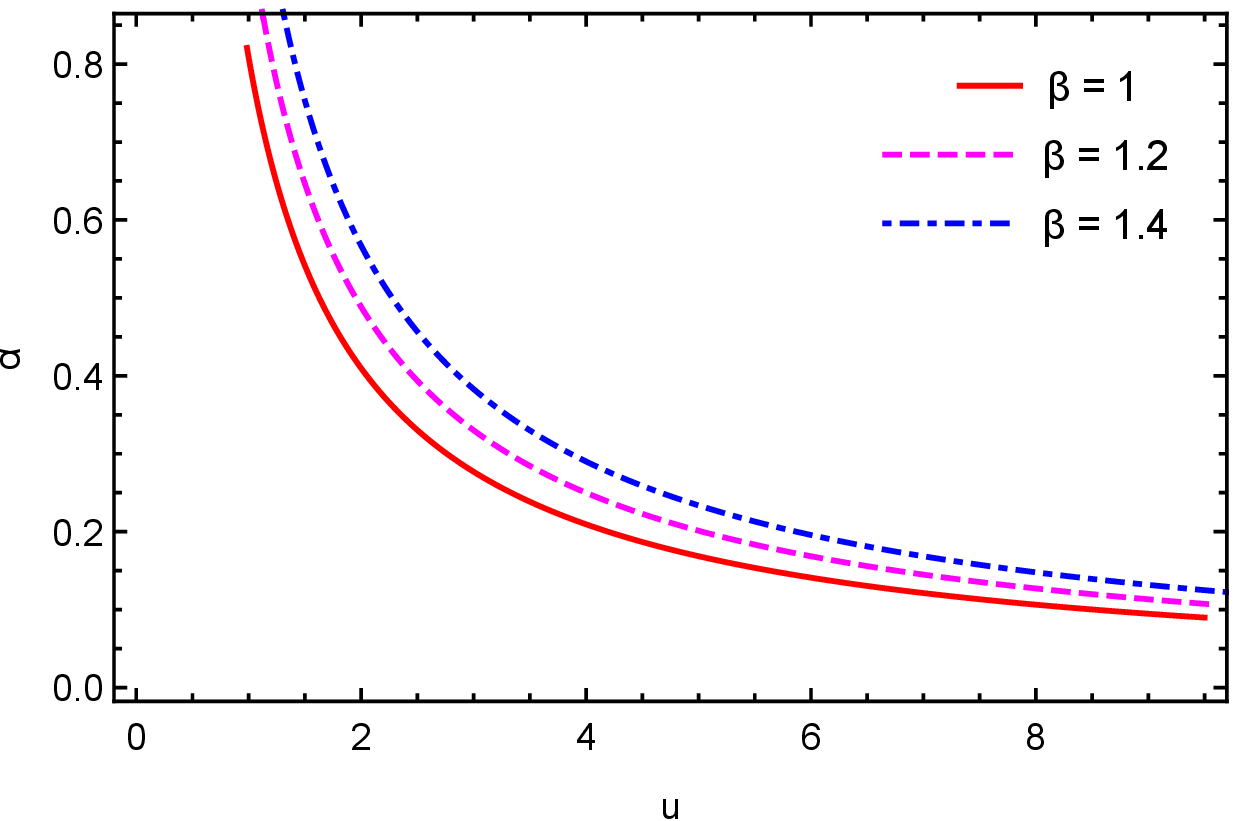}}\hspace{.3cm}
	\caption{Plots for effective potential and deflection angle for redshift function $\Phi_{eff}(r)=\frac{1}{2}\log_e(1-\frac{2\beta}{3r})$, where $\beta>r_0$.}
\end{figure}

\begin{figure}
	\centering
	\subfigure[]{\includegraphics[height=4.3cm, width=5cm]{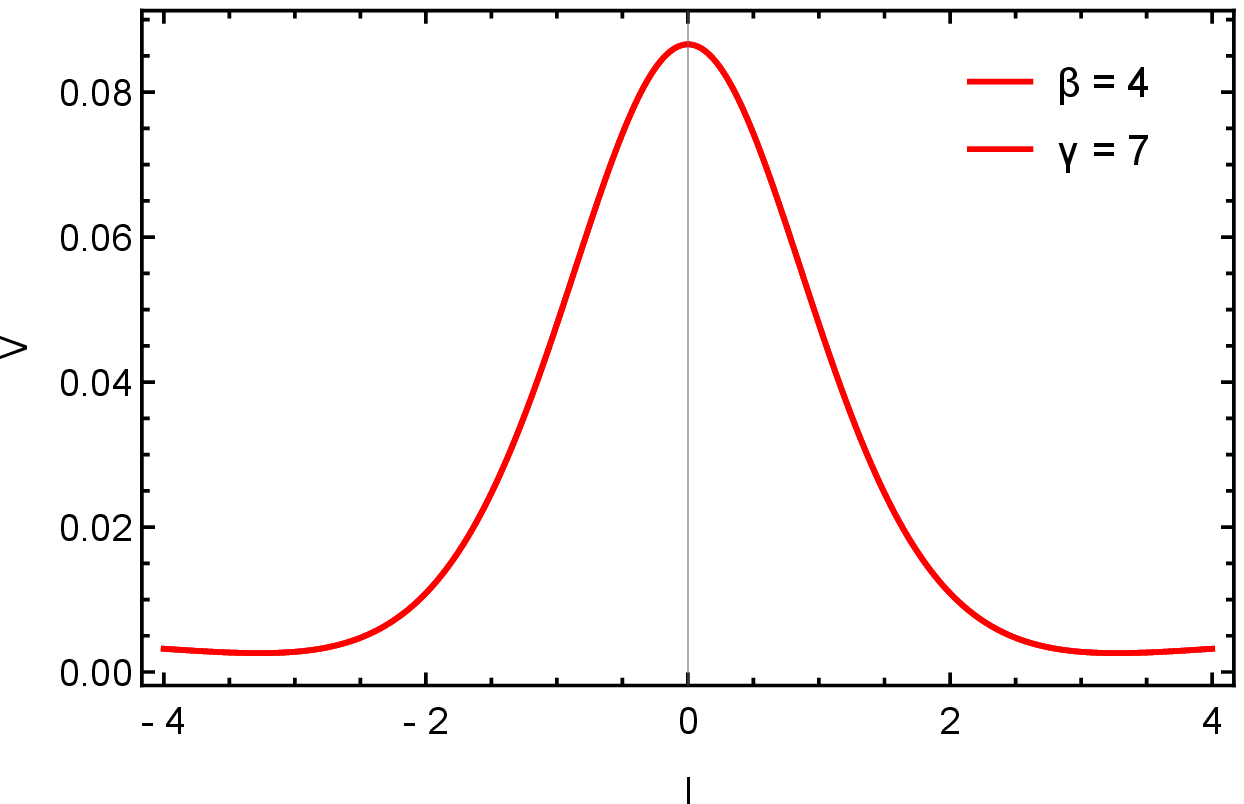}}\hspace{.5cm}
	\subfigure[]{\includegraphics[height=4.3cm, width=5cm]{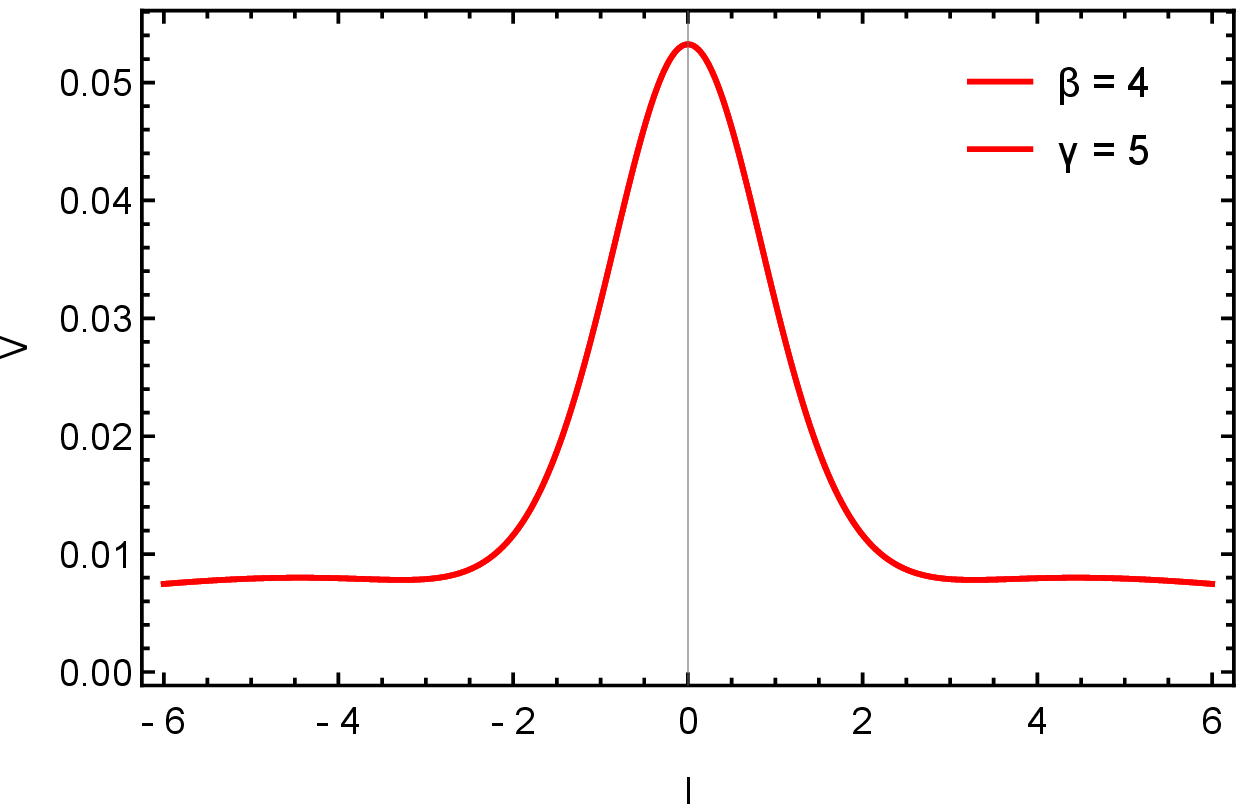}}\hspace{.5cm}
	\subfigure[]{\includegraphics[height=4.3cm, width=5cm]{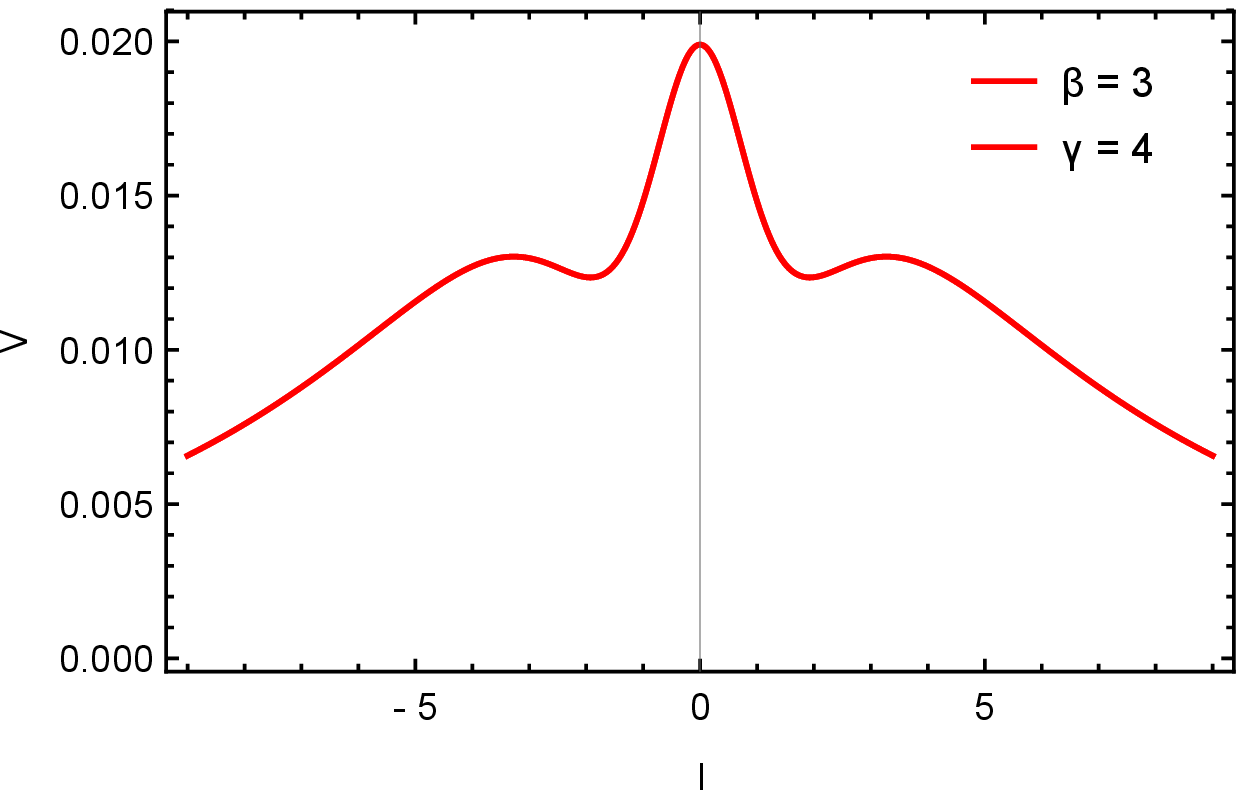}}\vspace{.5cm}
	\subfigure[]{\includegraphics[height=4.3cm, width=5cm]{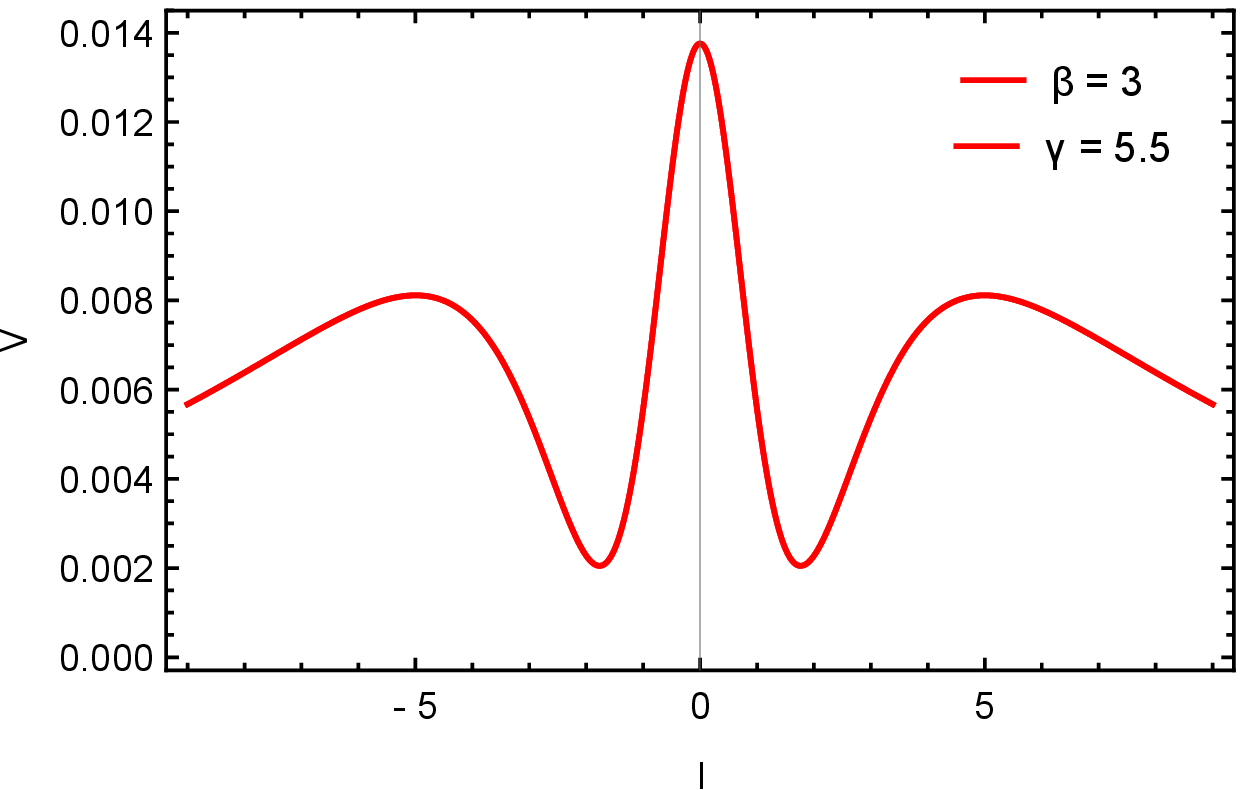}}\hspace{.5cm}
	\subfigure[]{\includegraphics[height=4.3cm, width=5cm]{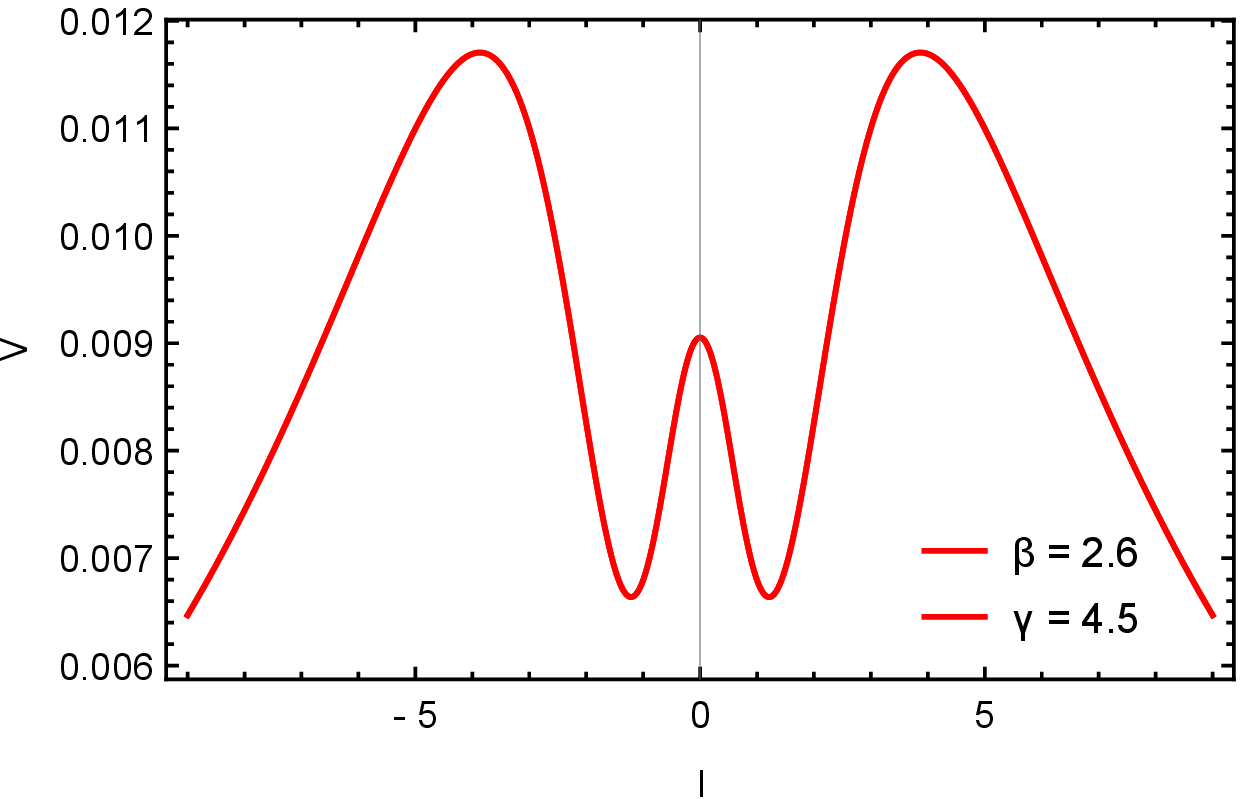}}\hspace{.5cm}
	\subfigure[]{\includegraphics[height=4.3cm, width=5cm]{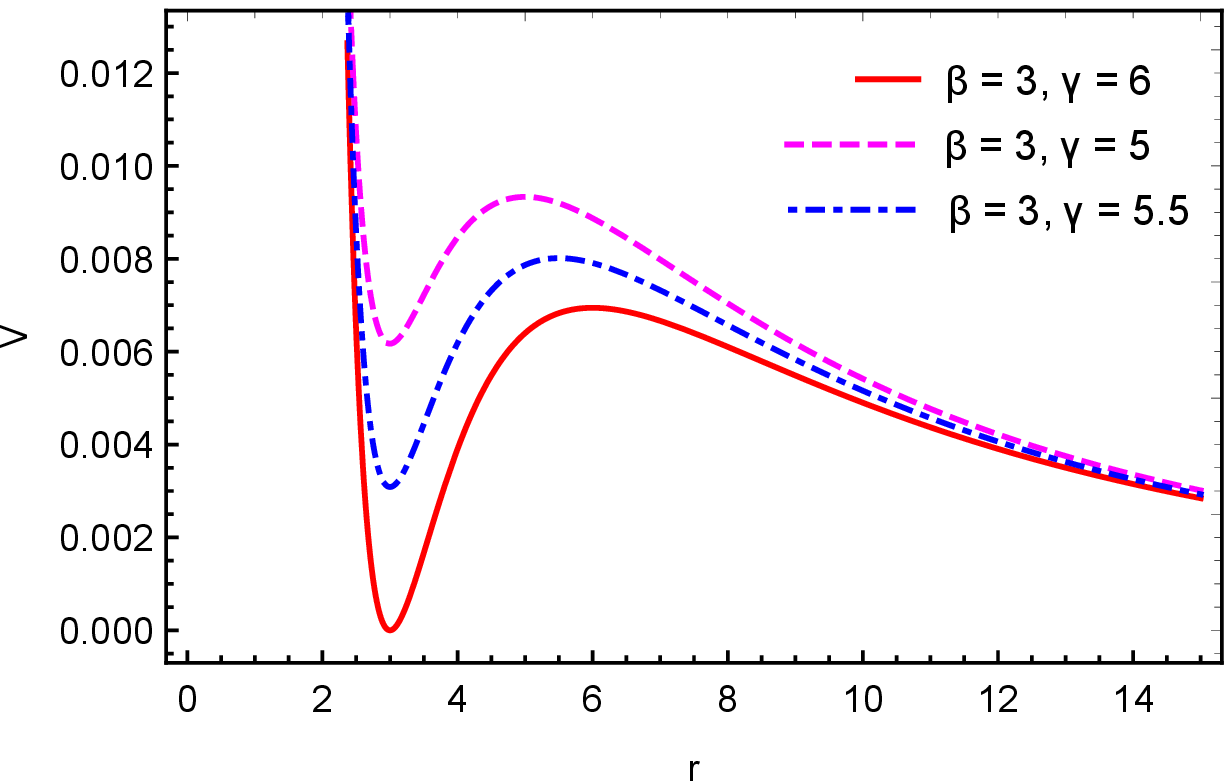}}\hspace{.5cm}
	\caption{Plots for effective potential for redshift function $\Phi_{eff}(r)=\frac{1}{2}\log_e\left(1-\frac{2(\beta+\gamma)}{3r}+\frac{\beta\gamma}{r^2}\right)$, where $r_0<\beta<\gamma$.}
\end{figure}

\begin{figure}
	\centering
	\subfigure[]{\includegraphics[height=4.3cm, width=5cm]{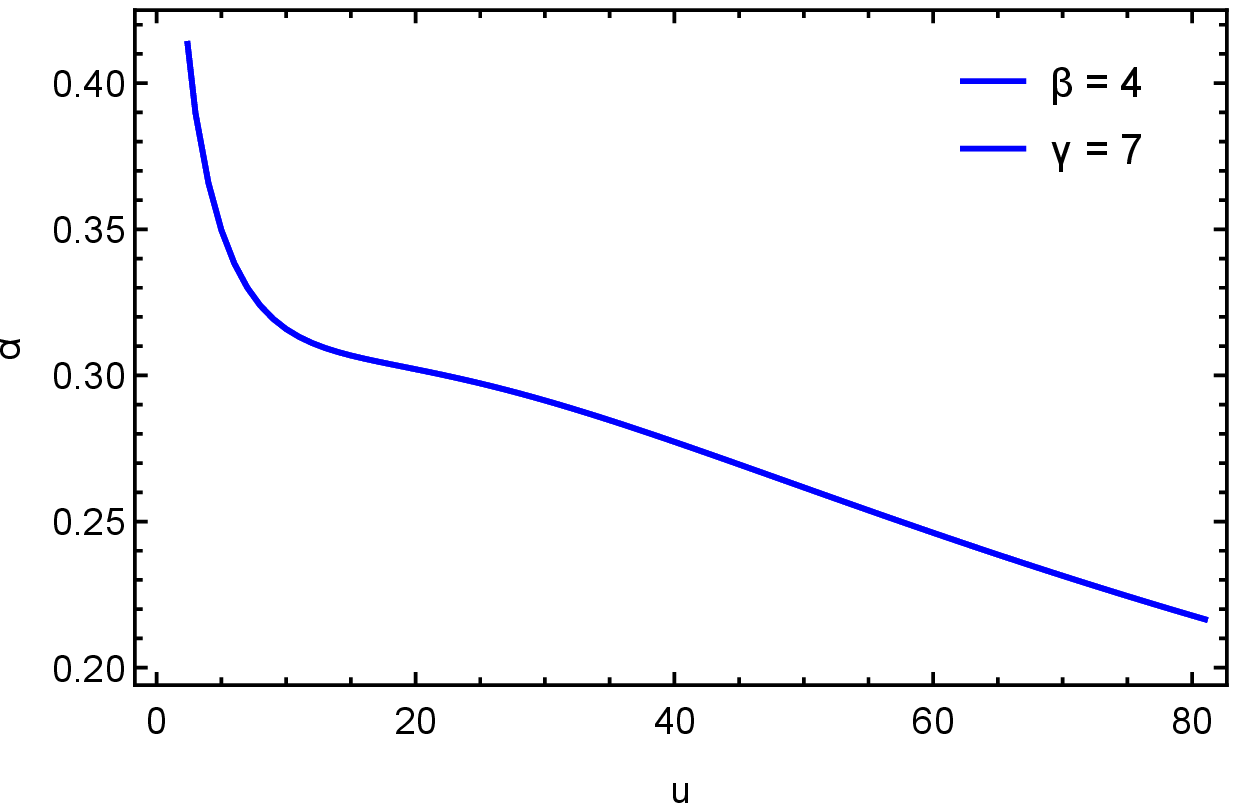}}\hspace{.5cm}
	\subfigure[]{\includegraphics[height=4.3cm, width=5cm]{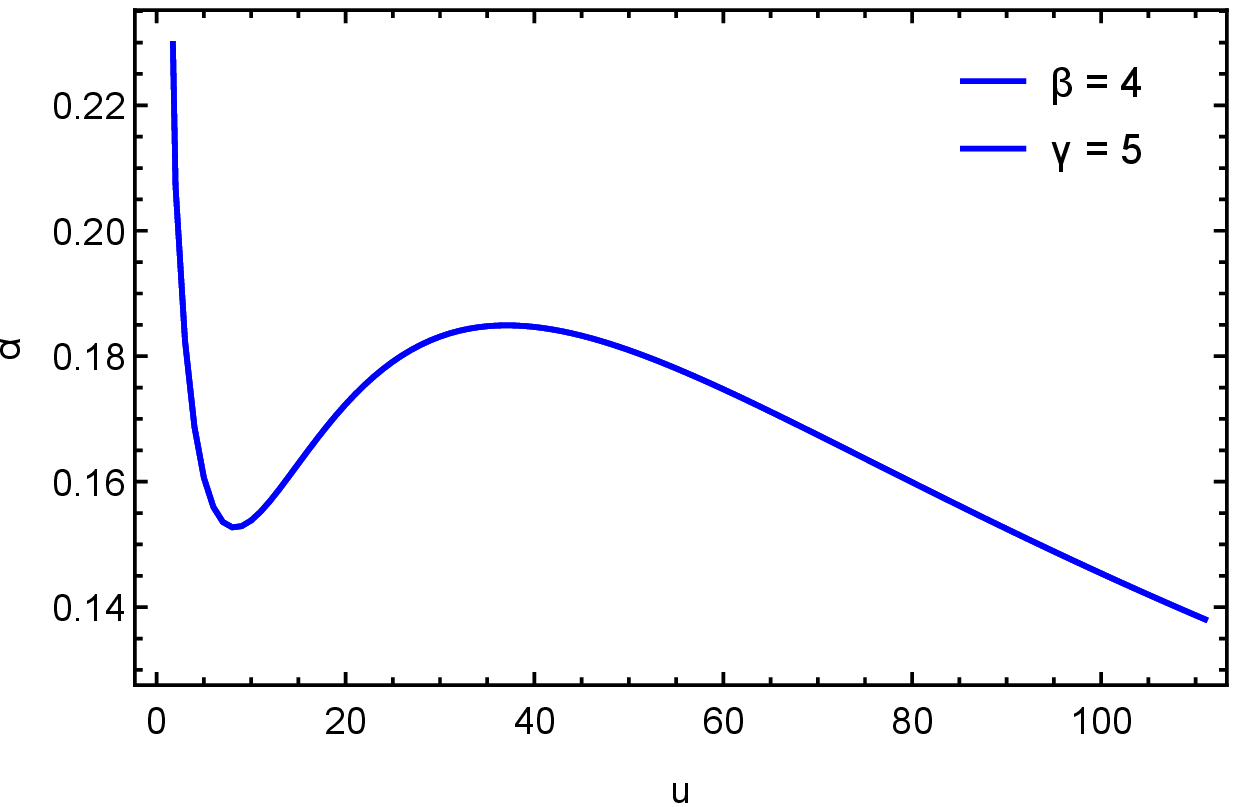}}\hspace{.5cm}
	\subfigure[]{\includegraphics[height=4.3cm, width=5cm]{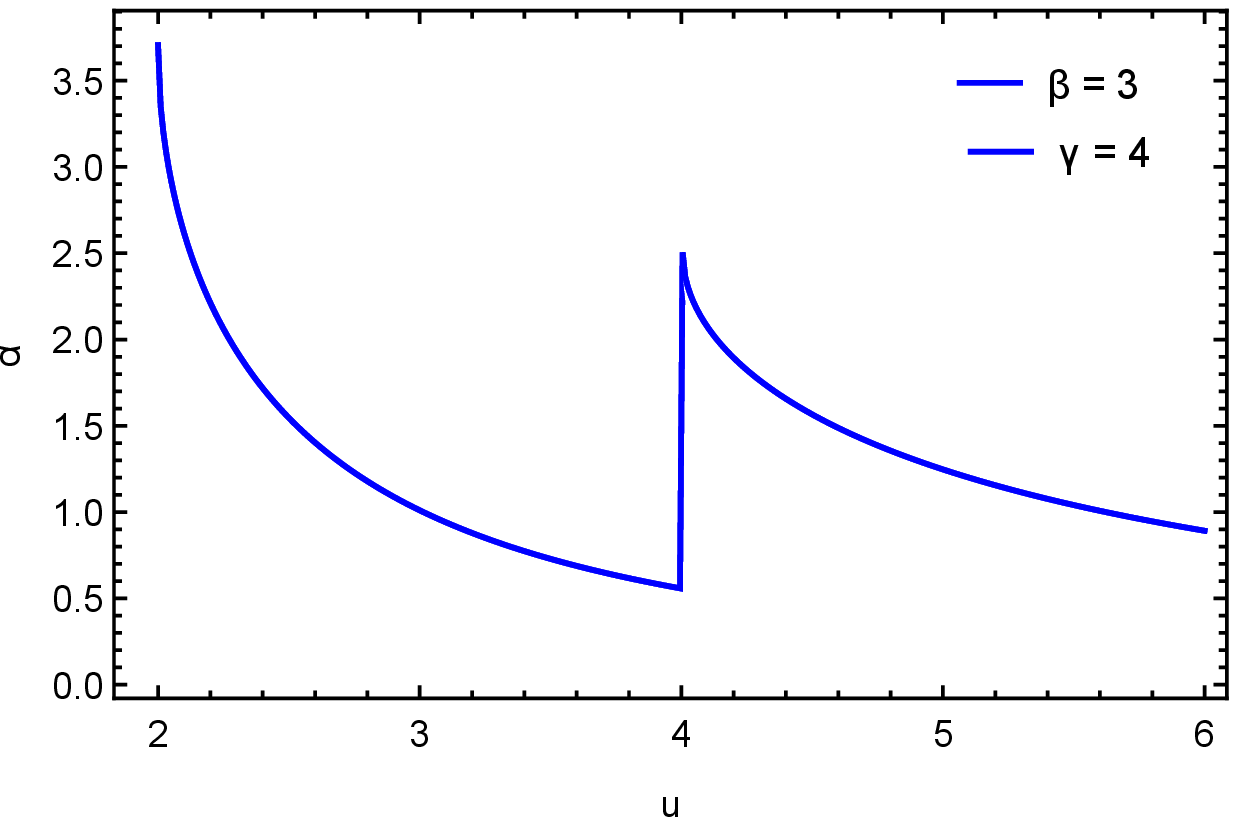}}\vspace{.5cm}
	\subfigure[]{\includegraphics[height=4.3cm, width=5cm]{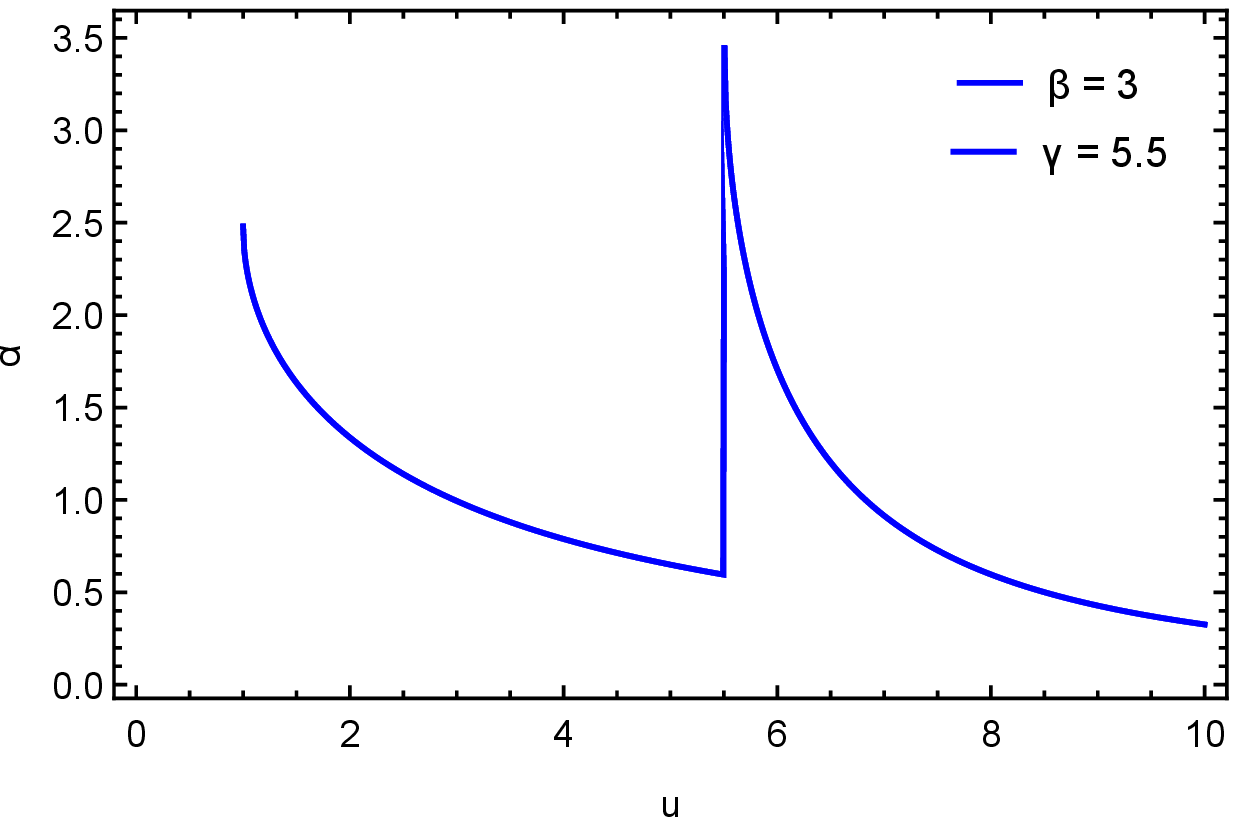}}\hspace{.5cm}
	\subfigure[]{\includegraphics[height=4.3cm, width=5cm]{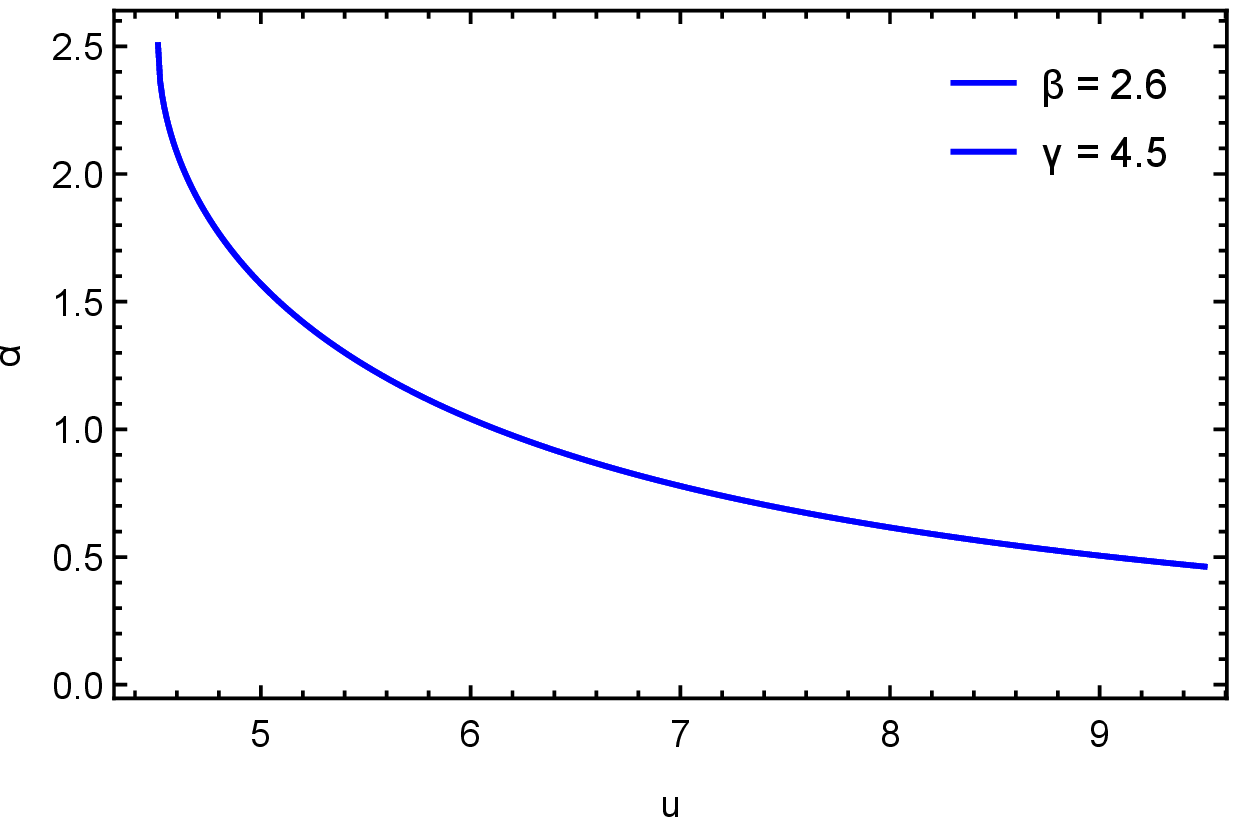}}\hspace{.5cm}
		\caption{Plots for deflection angle for redshift function $\Phi_{eff}(r)=\frac{1}{2}\log_e\left(1-\frac{2(\beta+\gamma)}{3r}+\frac{\beta\gamma}{r^2}\right)$, where $r_0<\beta<\gamma$.}
\end{figure}

\section{Energy conditions}
This section is focused on the exploration of energy conditions for charged wormholes with metric functions considered in previous section. The background of $f(R,T)$ theory of gravity proposed by Harko et al. \cite{Harko} with $f(R,T)=R+\lambda T$, where $\lambda$ is a constant, is adopted. Using this model with matter Lagrangian $L_m=p$ and metric \eqref{charge}, the field equations reduce to
\begin{equation}\label{fe}
R_{\mu}^{\nu} -\dfrac{1}{2}R\delta_{\mu}^{\nu}=8\pi( T_{\mu}^{\nu}+ T_{\mu}^{\nu e})+2\lambda T_{\mu}^{\nu} + (-2p+ T)\lambda\delta_{\mu}^{\nu},
\end{equation}
with electromagnetic energy momentum tensor $T_{\mu}^{\nu e}=\frac{q^2}{8\pi r^4}(-1,-1,1,1)$ and the energy momentum tensor for the matter source of the wormholes $T_{\mu}^{\nu} = (\rho + p_t)u_\mu u^\nu + p_t\delta_{\mu}^{\nu} +(p_r-p_t)X_\mu X^\nu
$, where $u^{\mu}u_\mu=-1$, $X^{\mu}X_\mu=1$, $p_r$ and $p_t$ are radial and tangential pressures respectively and $\rho$ stands for the energy density.
From \eqref{fe}, the field equations are obtained as

\begin{eqnarray}\label{fe1}
\frac{-{r_0}^2}{r^4}&=&(8\pi+\lambda)\rho-\lambda(p_r+2p_t),
\end{eqnarray}

\begin{eqnarray}\label{fe2}
&&\frac{1}{r^4 \left(r^2 e^{2 {\Phi_{eff}}(r)}+q^2\right)}\Big[2 r^3 e^{2 {\Phi_{eff}}(r)} {\Phi_{eff}}'(r) \left(q^2+r^2-{r_0}^2\right)+q^2 \left(r^2 \left(e^{2 {\Phi_{eff}}(r)}-2\right)+{r_0}^2\right)-r^2 {r_0}^2 e^{2 {\Phi_{eff}}(r)}\nonumber\\
&-&q^4+q^2\left(r^2 e^{2 {\Phi_{eff}}(r)}+q^2\right)\Big]=(8\pi+3\lambda)p_r+\lambda(\rho+2p_t),
\end{eqnarray}

\begin{eqnarray}\label{fe3}
&&\frac{1}{r^4 \left(r^2 e^{2 {\Phi_{eff}}(r)}+q^2\right)^2}\Bigg[r^4 e^{2 {\Phi_{eff}}(r)} \left(r^2 e^{2 {\Phi_{eff}}(r)}+q^2\right) {\Phi}''_{eff}(r) \left(q^2+r^2-{r_0}^2\right)+r^4 e^{2 {\Phi_{eff}}(r)} \left(r^2 e^{2 {\Phi_{eff}}(r)}+2 q^2\right) \nonumber\\
&\times&{\Phi}'_{eff}(r)^2 \left(q^2+r^2-{r_0}^2\right)+r^3 e^{2 {\Phi_{eff}}(r)} {\Phi}'_{eff}(r) \left(r^4 e^{2 {\Phi_{eff}}(r)}+2 q^4+q^2 \left(3 r^2-2 {r_0}^2\right)\right)+q^4 r^2 e^{2 {\Phi}(r)}\nonumber\\
&+&2 q^2 r^4 e^{2 {\Phi_{eff}}(r)}-q^2 r^4 e^{4 {\Phi_{eff}}(r)}-q^2 r^2 {r_0}^2 e^{2 {\Phi_{eff}}(r)}+r^4 {r_0}^2 e^{4 {\Phi_{eff}}(r)}+q^6+q^4 r^2-q^4 {r_0}^2+q^2 \nonumber\\
&\times&\left(r^2 e^{2 {\Phi_{eff}}(r)}+q^2\right)^2\Bigg]=(8\pi+4\lambda)p_t +(\rho+p_r)\lambda.
\end{eqnarray}
Solving Equations \eqref{fe1}-\eqref{fe3},
\begin{eqnarray}
\rho&=&\frac{1}{8 \pi ^2 (\lambda +4 \pi ) r^4 \left(r^2 e^{2 \phi_{eff}(r)}+q^2\right)^2}\Bigg[-\pi  \lambda  r^4 e^{2 \phi_{eff}(r)} \left(r^2 e^{2 \phi(r)}+q^2\right) \phi''(r) \left(q^2+r^2-r_0^2\right)-\pi  \lambda  r^4 e^{2 \phi(r)}\times\nonumber\\
&&\left(r^2 e^{2 \phi_{eff}(r)}+2 q^2\right) \phi'_{eff}(r)^2 \left(q^2+r^2-r_0^2\right)-\pi  \lambda  r^3 e^{2 \phi_{eff}(r)} \phi'_{eff}(r) \left(q^2 \left(r^2 \left(e^{2 \phi_{eff}(r)}+4\right)-3 r_0^2\right)+r^2 e^{2 \phi_{eff}(r)}\right.\times\nonumber\\
&& \left.\left(2 r^2-r_0^2\right)+3 q^4\right)+2 \lambda ^2 q^4 r^2 e^{2 \phi_{eff}(r)}+\lambda ^2 q^2 r^4 e^{4 \phi_{eff}(r)}-\pi  \lambda  q^2 r^4 e^{2 \phi_{eff}(r)}+\pi  \lambda  q^2 r^4 e^{4 \phi_{eff}(r)}+2 \pi  \lambda  q^2 r^2 r_0^2 e^{2 \phi_{eff}(r)}-\nonumber\\
&&8 \pi ^2 q^2 r^2 r_0^2 e^{2 \phi_{eff}(r)}-4 \pi ^2 r^4 r_0^2 e^{4 \phi_{eff}(r)}+\lambda ^2 q^6+\pi  \lambda  q^4 r_0^2-4 \pi ^2 q^4 r_0^2\Bigg]
\end{eqnarray}

\begin{eqnarray}
p_r&=&-\frac{1}{8 \pi ^2 (\lambda +4 \pi ) r^4 \left(r^2 e^{2 \phi_{eff}(r)}+q^2\right)^2}\Bigg[-\pi  \lambda  r^4 e^{2 \phi_{eff}(r)} \left(r^2 e^{2 \phi_{eff}(r)}+q^2\right) \phi''_{eff}(r) \left(q^2+r^2-r_0^2\right)-\pi  \lambda  r^4 e^{2 \phi_{eff}(r)}\times\nonumber\\
&& \left(r^2 e^{2 \phi_{eff}(r)}+2 q^2\right) \phi'_{eff}(r)^2 \left(q^2+r^2-r_0^2\right)-\pi  r^3 e^{2 \phi_{eff}(r)} \phi'_{eff}(r) \left(8 \pi  \left(r^2 e^{2 \phi_{eff}(r)}+q^2\right) \left(q^2+r^2-r_0^2\right)+\right.\nonumber\\
&&\left.\lambda  \left(q^2 \left(r^2 \left(e^{2 \phi_{eff}(r)}+4\right)-3 r_0^2\right)+r^2 e^{2 \phi_{eff}(r)} \left(2 r^2-r_0^2\right)+3 q^4\right)\right)+2 \lambda ^2 q^4 r^2 e^{2 \phi_{eff}(r)}-8 \pi ^2 q^4 r^2 e^{2 \phi_{eff}(r)}+\nonumber\\
&&\lambda ^2 q^2 r^4 e^{4 \phi_{eff}(r)}-\pi  \lambda  q^2 r^4 e^{2 \phi_{eff}(r)}+\pi  \lambda  q^2 r^4 e^{4 \phi_{eff}(r)}+8 \pi ^2 q^2 r^4 e^{2 \phi_{eff}(r)}-8 \pi ^2 q^2 r^4 e^{4 \phi_{eff}(r)}+2 \pi  \lambda  q^2 r^2 r_0^2 e^{2 \phi_{eff}(r)}+\nonumber\\
&&4 \pi ^2 r^4 r_0^2 e^{4 \phi_{eff}(r)}+\lambda ^2 q^6+8 \pi ^2 q^4 r^2+\pi  \lambda  q^4 r_0^2-4 \pi ^2 q^4 r_0^2\Bigg]
\end{eqnarray}

\begin{eqnarray}
p_t&=&\frac{1}{32 \pi ^3 (\lambda +4 \pi ) r^4 \left(r^2 e^{2 \phi_{eff}(r)}+q^2\right)^2}\Bigg[\pi  \left(-\lambda ^2+4 \pi  \lambda +16 \pi ^2\right) r^4 e^{2 \phi_{eff}(r)} \left(r^2 e^{2 \phi_{eff}(r)}+q^2\right) \phi''_{eff}(r) \times\nonumber\\
&&\left(q^2+r^2-r_0^2\right)+\pi  \left(-\lambda ^2+4 \pi  \lambda +16 \pi ^2\right) r^4 e^{2 \phi_{eff}(r)} \left(r^2 e^{2 \phi_{eff}(r)}+2 q^2\right) \phi'_{eff}(r)^2 \left(q^2+r^2-r_0^2\right)+\nonumber\\
&&\pi  r^3 e^{2 \phi_{eff}(r)} \phi'_{eff}(r) \left(\lambda ^2 \left(-\left(q^2 \left(r^2 \left(e^{2 \phi_{eff}(r)}+4\right)-3 r_0^2\right)+r^2 e^{2 \phi_{eff}(r)} \left(2 r^2-r_0^2\right)+3 q^4\right)\right)+4 \pi  \lambda\right. \times\nonumber\\
&&\left. \left(-q^2 \left(r^2 \left(e^{2 \phi_{eff}(r)}-2\right)+r_0^2\right)+r^2 r_0^2 e^{2 \phi_{eff}(r)}+q^4\right)+16 \pi ^2 \left(r^4 e^{2 \phi_{eff}(r)}+2 q^4+q^2 \left(3 r^2-2 r_0^2\right)\right)\right)+\nonumber\\
&&
2 \lambda ^3 q^4 r^2 e^{2 \phi_{eff}(r)}-8 \pi  \lambda ^2 q^4 r^2 e^{2 \phi_{eff}(r)}-40 \pi ^2 \lambda  q^4 r^2 e^{2 \phi_{eff}(r)}-16 \pi ^3 q^4 r^2 e^{2 \phi_{eff}(r)}+\lambda ^3 q^2 r^4 e^{4 \phi_{eff}(r)}-\nonumber\\
&&\pi  \lambda ^2 q^2 r^4 e^{2 \phi_{eff}(r)}-3 \pi  \lambda ^2 q^2 r^4 e^{4 \phi_{eff}(r)}+12 \pi ^2 \lambda  q^2 r^4 e^{2 \phi_{eff}(r)}-28 \pi ^2 \lambda  q^2 r^4 e^{4 \phi_{eff}(r)}+32 \pi ^3 q^2 r^4 e^{2 \phi_{eff}(r)}-\nonumber\\
&&32 \pi ^3 q^2 r^4 e^{4 \phi_{eff}(r)}+2 \pi  \lambda ^2 q^2 r^2 r_0^2 e^{2 \phi_{eff}(r)}-8 \pi ^2 \lambda  q^2 r^2 r_0^2 e^{2 \phi_{eff}(r)}-16 \pi ^3 q^2 r^2 r_0^2 e^{2 \phi_{eff}(r)}+4 \pi ^2 \lambda  r^4 r_0^2 e^{4 \phi_{eff}(r)}+\nonumber\\
&&
16 \pi ^3 r^4 r_0^2 e^{4 \phi_{eff}(r)}+\lambda ^3 q^6-4 \pi  \lambda ^2 q^6-16 \pi ^2 \lambda  q^6+8 \pi ^2 \lambda  q^4 r^2+16 \pi ^3 q^4 r^2+\pi  \lambda ^2 q^4 r_0^2-8 \pi ^2 \lambda  q^4 r_0^2-\nonumber\\
&&16 \pi ^3 q^4 r_0^2\Bigg]
\end{eqnarray}

\begin{eqnarray}
\rho+p_r&=&\frac{1}{4 \pi ^2 (\lambda +4 \pi ) r^4 \left(r^2 e^{2 \phi_{eff}(r)}+q^2\right)^2}\Bigg[\pi  \lambda  r^4 e^{2 \phi_{eff}(r)} \left(r^2 e^{2 \phi_{eff}(r)}+q^2\right) \phi''_{eff}(r) \left(q^2+r^2-r_0^2\right)+\pi  \lambda  r^4 e^{2 \phi_{eff}(r)}\times\nonumber\\
&&
 \left(r^2 e^{2 \phi_{eff}(r)}+2 q^2\right) \phi'_{eff}(r)^2 \left(q^2+r^2-r_0^2\right)+\pi  r^3 e^{2 \phi_{eff}(r)} \phi'_{eff}(r) \left(4 \pi  \left(r^2 e^{2 \phi_{eff}(r)}+q^2\right) \left(q^2+r^2-r_0^2\right)+\right.\nonumber\\
 &&\left.\lambda  \left(q^2 \left(r^2 \left(e^{2 \phi_{eff}(r)}+4\right)-3 r_0^2\right)+r^2 e^{2 \phi_{eff}(r)} \left(2 r^2-r_0^2\right)+3 q^4\right)\right)-q^2 \left(\lambda ^2 \left(r^2 e^{2 \phi_{eff}(r)}+q^2\right)^2-\right.\nonumber\\
 &&\left.4 \pi ^2 \left(r^2 e^{2 \phi_{eff}(r)}+q^2\right) \left(r^2 \left(e^{2 \phi_{eff}(r)}-1\right)+r_0^2\right)+\pi  \lambda  \left(r^4 e^{4 \phi_{eff}(r)}-r^2 e^{2 \phi_{eff}(r)} \left(r^2-2 r_0^2\right)+q^2 r_0^2\right)\right)\Bigg]
\end{eqnarray}

\begin{eqnarray}
\rho+p_t&=&\frac{1}{32 \pi ^3 (\lambda +4 \pi ) r^4 \left(r^2 e^{2 \phi_{eff}(r)}+q^2\right)^2}\Bigg[\pi  \left(16 \pi ^2-\lambda ^2\right) r^4 e^{2 \phi_{eff}(r)} \left(r^2 e^{2 \phi_{eff}(r)}+q^2\right) \phi''_{eff}(r) \left(q^2+r^2-r_0^2\right)+\nonumber\\
&&
\pi  \left(16 \pi ^2-\lambda ^2\right) r^4 e^{2 \phi_{eff}(r)} \left(r^2 e^{2 \phi_{eff}(r)}+2 q^2\right) \phi'_{eff}(r)^2 \left(q^2+r^2-r_0^2\right)+\pi  r^3 e^{2 \phi_{eff}(r)} \phi'_{eff}(r) \left(-8 \pi  \lambda\right. \times\nonumber\\
&&\left. \left(r^2 e^{2 \phi_{eff}(r)}+q^2\right) \left(q^2+r^2-r_0^2\right)+\lambda ^2 \left(-\left(q^2 \left(r^2 \left(e^{2 \phi_{eff}(r)}+4\right)-3 r_0^2\right)+r^2 e^{2 \phi_{eff}(r)} \left(2 r^2-r_0^2\right)+\right.\right.\right.\nonumber\\
&&
\left.\left.\left.3 q^4\right)\right)+16 \pi ^2 \left(r^4 e^{2 \phi_{eff}(r)}+2 q^4+q^2 \left(3 r^2-2 r_0^2\right)\right)\right)+2 \lambda ^3 q^4 r^2 e^{2 \phi_{eff}(r)}-40 \pi ^2 \lambda  q^4 r^2 e^{2 \phi_{eff}(r)}-\nonumber\\
&&16 \pi ^3 q^4 r^2 e^{2 \phi_{eff}(r)}+\lambda ^3 q^2 r^4 e^{4 \phi_{eff}(r)}-\pi  \lambda ^2 q^2 r^4 e^{2 \phi_{eff}(r)}+\pi  \lambda ^2 q^2 r^4 e^{4 \phi_{eff}(r)}+8 \pi ^2 \lambda  q^2 r^4 e^{2 \phi_{eff}(r)}-24 \pi ^2 \lambda\times\nonumber\\
&&  q^2 r^4 e^{4 \phi_{eff}(r)}+32 \pi ^3 q^2 r^4 e^{2 \phi_{eff}(r)}-32 \pi ^3 q^2 r^4 e^{4 \phi_{eff}(r)}+2 \pi  \lambda ^2 q^2 r^2 r_0^2 e^{2 \phi_{eff}(r)}-48 \pi ^3 q^2 r^2 r_0^2 e^{2 \phi_{eff}(r)}+\nonumber\\
&&4 \pi ^2 \lambda  r^4 r_0^2 e^{4 \phi_{eff}(r)}+\lambda ^3 q^6-16 \pi ^2 \lambda  q^6+8 \pi ^2 \lambda  q^4 r^2+16 \pi ^3 q^4 r^2+\pi  \lambda ^2 q^4 r_0^2-4 \pi ^2 \lambda  q^4 r_0^2-\nonumber\\
&& 32 \pi ^3 q^4 r_0^2\Bigg]
\end{eqnarray}

From Equations \eqref{fe1}-\eqref{fe3}, we have found $\rho$, $p_r$ and $p_t$, and examined the null and weak energy conditions. The null energy condition (NEC) is defined as
$NEC\Leftrightarrow T_{\mu\nu}k^{\mu}k^{\nu}\ge 0$, for any null vector $k^{\mu}$. In terms of the principal pressures, it is defined as $NEC\Leftrightarrow ~~ \forall i, ~\rho+p_{i}\ge 0$. The weak energy condition (WEC) is defined as $WEC\Leftrightarrow T_{\mu\nu}V^{\mu}V^{\nu}\ge 0$, for a time-like vector $V^{\mu}$. In terms of the principal pressures, it is defined as $WEC\Leftrightarrow \rho\ge 0;$ and $\forall i,  ~~ \rho+p_{i}\ge 0$. For wormhole modelling, these energy conditions reduce to: NEC $\Leftrightarrow$ $\rho+p_r\geq 0$, $\rho+p_t\geq 0$; WEC $\Leftrightarrow$ $\rho\geq 0$, $\rho+p_r\geq 0$, $\rho+p_t\geq 0$. These ECs are tested in three cases corresponding to three forms of $\Phi(r)$: Case 1. $\Phi_{eff}(r)=\frac{1}{2}\log_e(1+\frac{2r_0}{3r})$, Case 2. $\Phi_{eff}(r)=\frac{1}{2}\log_e(1-\frac{2\beta}{3r})$, where $\beta>r_0$ and
Case 3. $\Phi_{eff}(r)=\frac{1}{2}\log_e\left(1-\frac{2(\beta+\gamma)}{3r}+\frac{\beta\gamma}{r^2}\right)$, where $r_0<\beta<\gamma$. In Case 1, both $\rho$ and $\rho+p_t$ are positive for every $r\geq r_0$ and $\lambda\neq -4\pi$. The other NEC term $\rho+p_r$ is positive for $\lambda<-4\pi$ and negative for  $\lambda>-4\pi$. Thus, WEC is satisfied for $r\geq r_0$ and $\lambda < -4\pi$.
In Case 2,  $\rho$ is positive for every $r\geq r_0$ and $\lambda\neq -4\pi$. The  NEC terms $\rho+p_r$ and $\rho+p_t$ are  positive for $\lambda<-4\pi$ and  $r\geq \beta$. Subsequently, if we put $\lambda=0$, then the gravitational field equations will be reduced to general Einstein theory, and coupled with electromagnetic field, in this situation NEC is violated, so exotic matter is required to sustain a wormhole solution in general relativity. Furthermore, if we avoid electromagnetic field, then it is observed that the NEC is violated in all the above cases, which indicates the presence of exotic matter near the throat. Eventually, for this study we conclude that modified gravity coupled with electromagnetic field could avoid the necessity of exotic matter for the construction of wormhole solutions. Thus, WEC is satisfied for $r\geq \beta$ and $\lambda < -4\pi$. Results of Case 3 are similar to Case 1. Hence, in Cases 1 and 3, wormholes are completely filled with non-exotic matter. In Case 2, after the occurrence of photon sphere, charged wormhole geometries are filled with non-exotic matter. This shows that the positive deflection angle corresponds to charged wormholes threaded with non-exotic matter.
\section{Conclusion}
This paper is focused on the study of strong gravitational lensing effect on charged wormholes. We have obtained the conditions for the existence of the photon spheres at and outside the throat in terms of effective potential $V_{eff}$. Then we have explained the existence of photon spheres  with the shape function $b(r)=\frac{r_0^2}{r}$ in  following three cases of redshift function $\Phi_{eff}(r)$:  Case 1. $\Phi_{eff}(r)=\frac{1}{2}\log_e(1+\frac{2r_0}{3r})$,  Case 2. $\Phi_{eff}(r)=\frac{1}{2}\log_e(1-\frac{2\beta}{3r})$, where $\beta>r_0$ and
Case 3. $\Phi_{eff}(r)=\frac{1}{2}\log_e\left(1-\frac{2(\beta+\gamma)}{3r}+\frac{\beta\gamma}{r^2}\right)$, where $r_0<\beta<\gamma$. In Case 1, $V_{eff}$ is maximum at the throat and deflection angle diverges at the throat and after that it decreases. This depicts the existence of photon sphere at the throat of wormhole. In Case 2, $V_{eff}$ is maximum  and  deflection angle diverges at   $r=\beta>r_0$. The point   $r=\beta$ corresponds to the radius of the photon sphere occurring outside the throat. Further, in Case 3, $V_{eff}$ possesses maximum value at  $r=r_0$ and  $r=\gamma>r_0$. Between two maxima, a minima occurs at $r=\beta$. This implies the existence of effective photon spheres at and outside the throat. The value  of $V_{eff}$ at the throat is greater than   its value at $r=\gamma$ which means that the throat will participate in the formation of relativistic images due to strong gravitational lensing.  The deflection angle diverges at  two points corresponding to the radii of the throat and outer photon sphere. In Cases 1 and 3, there will be a formation of only one set of closely packed Einstein's relativistic rings at and outside the throat respectively. However, in Case 2,  there will be a formation of two sets of Einstein's relativistic rings at and outside the throat in which the set of rings formed at the throat would be closely packed in comparison of other set of rings.   Furthermore, we have investigated the matter threading the wormhole structure. For this purpose, we have derived the field equations in the context of $f(R,T)=R+\lambda T$ gravity model and tested  null and weak energy conditions. In Cases 1 and 3, these energy conditions are satisfied for $r\geq r_0$ with $\lambda<-4\pi$. In Case 2,  ECs are valid for $r\geq\beta$ with $\lambda<-4\pi$. This implies  the existence of non-exotic matter at and outside the (i) effective photon sphere existing at the throat in Cases 1 and 3; (ii) outer photon sphere in Case 2. Thus, we have explored the photon spheres due to strong gravitational lensing and nature of matter inside the charged wormhole geometry which may be helpful in the investigation of wormholes in future observations.\\

{\bf Acknowledgment:} The authors are very much thankful to the reviewer and editor for their
 constructive comments for the improvement of the paper.

\end{document}